\documentclass{article}

\usepackage[english]{babel}

\usepackage[a4paper,top=2cm,bottom=2cm,left=3cm,right=3cm,marginparwidth=1.75cm]{geometry}

\usepackage{amsmath}
\usepackage{amssymb}
\usepackage{bm}

\usepackage{graphicx}
\usepackage{subcaption}
\usepackage{booktabs}
\usepackage{multirow}
\usepackage{array}
\usepackage[colorlinks=true, allcolors=blue]{hyperref}
\usepackage{lineno} % Line numbers
\usepackage{authblk}

\usepackage{parskip}
\usepackage{algorithm}
\usepackage{algorithmic}

\graphicspath{{figures/}}

\usepackage[numbers,sort&compress]{natbib}

\usepackage{xcolor}

\newcommand{\lesets}{GNN-oracle}

\newcommand{\bvae}{bVAE}
\newcommand{\binaryfm}{BinaryFM}

\newcommand{\zbin}{\bm{z}}
\newcommand{\xcomp}{\bm{x}}

\newcommand{\oracle}{\hat{y}_{\mathrm{GNN}}}

\newcommand{\argmax}{\operatorname*{arg\,max}}
\newcommand{\argmin}{\operatorname*{arg\,min}}

\newcommand{\randomcomp}{Random (comp)}
\newcommand{\gacomp}{GA (comp)}
\newcommand{\rfucbcomp}{RF-UCB (comp)}
\newcommand{\randomlatent}{Random (latent)}
\newcommand{\randompertlatent}{Random+Pert (latent)}
\newcommand{\galatent}{GA (latent)}
\newcommand{\workflowpert}{Workflow}
\newcommand{\workflownopert}{Workflow (no pert)}

\title{QUBO-Compatible Active Learning for Inverse Design of High-Entropy Alloys}

\author[1]{Giorgio Silvi} \author[1]{Kirsten Bark} \author[1]{Rolando Reiner} \author[1]{Nicolas Vogt} \author[2]{Thomas Plehn} \author[2]{Daniel Barragan-Yani} \author[2]{Marc Landmann} \author[2]{David Melching} \affil[1]{HQS Quantum Simulations GmbH, 76131 Karlsruhe, Germany} \affil[2]{Institute for Frontier Materials on Earth and in Space, German Aerospace Center (DLR), Linder Hoehe, Cologne, 51147, Germany}

\date{August 28, 2026}

\begin{document}
\maketitle

\begin{abstract}
Machine-learned forward models can rapidly predict alloy properties, but
their use for inverse design remains challenging when the search should also
retain compatibility with quadratic unconstrained binary optimization (QUBO).
Here, we develop a QUBO-compatible active-learning framework for inverse design
of high-entropy alloys using a pretrained graph-neural-network predictor as a
fixed property oracle. A property-guided binary variational autoencoder provides
a binary latent representation, while an ensemble of quadratic factorization
machines guides candidate selection.
We systematically benchmark the framework through controlled latent-space
ablations and comparison with direct composition-space optimization. The results
show that candidate generation is a major determinant of search performance:
local perturbations around previously high-performing latent codes provide the
largest workflow-specific improvement, while surrogate-based selection further
prioritizes candidates within the enriched search pool. The resulting
QUBO-compatible workflow remains competitive with strong classical optimization
strategies, although a composition-space genetic algorithm achieves the highest
mean score. Finally, the learned quadratic surrogate can be exported directly
as a QUBO. These results show that effective data acquisition can be separated
from the final QUBO optimization endpoint, providing a benchmarked route for
QUBO-compatible data-driven materials inverse design.
\end{abstract}

\section{Introduction}
\label{sec:introduction}

% HEAs and inverse-design problem
High-entropy alloys (HEAs) and related multi-principal-element alloys offer a large compositional design space in which several elements can be combined to tune mechanical, thermal, chemical, and functional properties \citep{George2019,Wen2019,Liu2023}. This flexibility turns materials discovery into a challenging undertaking: even when the number of constituent elements is restricted, the number of possible compositions grows rapidly, while accurate evaluation by atomistic simulation or experiment remains comparatively expensive. Machine-learning approaches have therefore become increasingly important for accelerating HEA design, including property prediction, high-throughput screening, active learning, and inverse design \citep{butler2018machine,rao2022machine,zeng2022machine,Zhichao2024}. Instead of exhaustively evaluating the design space, these approaches aim to learn from previously available data and concentrate new evaluations on promising candidate materials.

% From forward prediction to inverse design
A central ingredient of such workflows is an accurate forward model that maps a candidate material to its predicted properties. Deep learning has become particularly effective for materials-property prediction, with graph neural networks providing flexible representations of chemical environments and atomic structure \citep{reiser2022graph}. For HEAs, permutation-invariant composition models \citep{zhang2022composition} and graph-based representations of local chemical environments \citep{zhang2025graph} have demonstrated that complex composition--property relationships can be learned from available data. However, a forward predictor alone does not solve the inverse-design problem: once rapid property evaluation is available, an optimization strategy is still required to determine which candidate should be proposed next. Surrogate-based optimization and active learning address this problem by iteratively using previous evaluations to guide the search toward promising regions \citep{lookman2019active,Kavalsky2023}, while generative models can provide lower-dimensional representations in which candidate materials are generated and optimized \citep{sanchezlengeling2018inverse,lu2022inverse,wang2020deep}.

% Data-driven QUBO optimization
Further challenges arise from the connection of data-driven inverse design with binary quadratic optimization. Many materials-design problems are inherently combinatorial because candidate materials are constructed from discrete choices of constituents, phases, morphologies, or atomic configurations. Quadratic unconstrained binary optimization (QUBO), or equivalently the Ising formulation, provides a common problem form for a wide range of classical, quantum-inspired and quantum optimization backends \citep{Lucas2014,Yarkoni2022}. However, even when a materials-design problem contains discrete or combinatorial variables, its objective function is rarely available directly in analytical QUBO form. In data-driven materials optimization, the relation between input design and target properties is typically only available through simulations, experiments, or predictive models. The optimization objective must therefore itself be learned from data. Factorization machines (FMs) are particularly suitable for this task, since their second-order prediction function becomes a quadratic polynomial for binary input variables and consequently can be expressed as a QUBO \citep{Rendle2010,kitai2020designing}. Data-driven QUBO approaches have subsequently been extended to continuous black-box optimization and increasingly complex materials-design problems \citep{Izawa2022,Plehn2025}.

% FM+QO
This is the central idea of factorization-machine-based QUBO optimization (FM+QO), in which an FM surrogate is trained on previously evaluated designs and the resulting quadratic model is optimized using a compatible QUBO backend. Depending on the application, this backend can be simulated annealing (SA) \citep{Kirkpatrick1983}, quantum annealing (QA) \citep{Kadowaki1998}, digital annealing (DA) \citep{Aramon2019}, or gate-based quantum optimization algorithms such as the quantum approximate optimization algorithm (QAOA) \citep{Farhi2014}. Selected candidates proposed by the FM model are evaluated, and used to update the surrogate \citep{kitai2020designing,Tamura2026,Plehn2025}. Related approaches have been applied to metamaterials and nanostructures, molecular design, and other materials-optimization problems \citep{wilson2021machine,Izawa2022,mao2023chemical,Xu2025,Hama2026}. Recent work has further extended data-driven QUBO optimization to multi-objective alloy-design and recycling problems \citep{Plehn2025,Plehn2026}, while quantum-annealing-assisted methods have been investigated specifically for HEAs, including NbMoTaW lattice optimization and quantum-assisted machine learning for HEA discovery \citep{Xu2025,IbarraHoyos2026}. These developments illustrate the potential of data-driven QUBO models as an interface between black-box materials objectives and binary optimization backends. Additionally, they expose an important representation problem: alloy compositions are continuous and constrained by normalization, positivity, and compositional-support requirements, and are therefore not naturally expressed as unconstrained binary variables.

% Binary representation learning & search problem
Binary representation learning provides a route to bridge this gap. Instead of directly discretizing each physical design variable, which can require many binary variables and can strongly affect the resulting surrogate landscape \citep{Endo2025}, a machine-learning model learns a discrete latent representation tailored to the structure of the underlying data. Candidate materials can then be reconstructed from this lower-dimensional binary space. Binary variational autoencoders have previously been combined with FMs and Ising/QUBO optimization to treat constrained continuous and molecular design spaces \citep{wilson2021machine,mao2023chemical}. Related work has also explored alternative binary encodings for continuous black-box optimization \citep{Izawa2022} and modular inverse-design architectures that separate learned representations from QUBO-based optimization \citep{Deguchi2026}. Such representations create finite binary search spaces on which quadratic surrogates can be defined directly, but they do not by themselves determine how these spaces should be searched. Recent developments in factorization-machine annealing likewise highlight the importance of balancing exploration and exploitation during black-box optimization \citep{Hama2026}. Optimization performance can therefore depend on how candidate latent codes are generated, how exploration and exploitation are balanced, and how effectively a surrogate prioritizes candidates from the available search pool. This motivates a systematic separation of the structural benefit of QUBO compatibility from the mechanisms that actually drive search performance within the binary representation.

% This paper
In this work, we study these questions for the problem of inverse-designing quaternary HEAs with high bulk moduli. We use the pretrained graph neural network (GNN) model \textsc{LESets} from Zhang et al. \cite{zhang2025graph} as a fixed forward prediction oracle. A property-guided bVAE maps the GNN representation of an alloy to binary latent codes and decodes them back to valid quaternary compositions. In this binary space, an ensemble of quadratic factorization machines is trained on latent code--property pairs. New candidates are selected using an upper-confidence-bound acquisition function over a finite candidate pool that combines broad latent sampling with local bit-flip perturbations of previously high-performing codes. The final learned quadratic surrogate is in QUBO form and can be optimized over the binary domain using a compatible backend.
Beyond introducing this workflow, a central objective of this study is to provide a systematic benchmark of the factors that determine optimization performance. Within the same pretrained binary representation, we compare random search, evolutionary optimization, perturbation-based local search, and surrogate-guided active learning, which includes direct ablations of the candidate-pool construction and investigation into the role of repeated QUBO optimization. We additionally compare these latent-space approaches with direct composition-space optimization under matched additional oracle-evaluation budgets. The resulting benchmark allows us to separate, to the extent it is possible, the effects of representation, candidate generation, and model-based selection.

% Structure
The remainder of the paper is organized as follows. 
Section~\ref{sec:problem-setting} defines the HEA inverse-design problem setup. 
Section~\ref{sec:workflow} introduces the binary latent representation, active-learning procedure, candidate-pool construction, and QUBO formulation. 
Section~\ref{sec:benchmark-methods} describes the latent- and composi-tion-space benchmark methods and evaluation protocol.
Section~\ref{sec:results} presents the benchmark results and workflow ablations, followed by their interpretation and limitations in
Sec.~\ref{sec:discussion}. 
Finally, Sec.~\ref{sec:conclusion} summarizes the main conclusions.
\section{Inverse-design problem}
\label{sec:problem-setting}

Our goal is to identify quaternary HEA compositions with high bulk modulus within a fixed pool of candidate elements. From a materials-design perspective, this is an inverse-design problem: rather than predicting the property of a prescribed alloy, we seek compositions that maximize a target property subject to a specified compositional domain. Direct evaluation of large numbers of candidate alloys using density functional theory (DFT) or experiment is resource intensive, motivating machine-learning approaches to materials design, including property prediction~\citep{butler2018machine}, generative composition design~\citep{sanchezlengeling2018inverse,lu2022inverse}, and surrogate-guided optimization~\citep{lookman2019active}. Machine-learning-based inverse design has also been applied specifically to high-entropy alloys~\citep{rao2022machine,zeng2022machine}. In the present work, we formulate the inverse-design task with respect to a pretrained and frozen property predictor, which serves as a common learned objective for evaluating candidate compositions.

Let $\mathcal{E}$ denote the fixed element vocabulary
$
\{
\mathrm{Al},\mathrm{Co},\mathrm{Cr},\mathrm{Cu},\mathrm{Fe},
\mathrm{Hf},\mathrm{Mn},\mathrm{Mo},\mathrm{Nb},\mathrm{Ni},
\mathrm{Ta},\mathrm{Ti},\mathrm{V},\mathrm{W},\mathrm{Zr}
\}
$
\cite{zhang2025graph} and let $\mathcal{X}$ denote the corresponding space of valid quaternary compositions. A composition $\xcomp$ is represented by its elemental atomic fractions and is required to satisfy non-negativity, normalization, and a support-cardinality constraint requiring exactly four nonzero components:
\begin{equation}
\mathcal{X}
=
\left\{
\xcomp \in \mathbb{R}_{\geq 0}^{|\mathcal{E}|}
:
\sum_{e \in \mathcal{E}} x_e = 1,
\;
\|\xcomp\|_0 = 4
\right\}.
\label{eq:composition-space}
\end{equation}
Thus, each candidate contains exactly four elements with nonzero atomic fractions. Equation~\ref{eq:composition-space} defines the common quaternary composition domain used throughout the benchmark; no additional lower or upper bounds on the four nonzero fractions are imposed. Details of the reference dataset and the oracle architecture are given in the supplementary information (Sec.~\ref{sec:supp-oracle}).

We use the pretrained GNN model \textsc{LESets} from \citep{zhang2025graph} as the property oracle and keep all of its parameters fixed throughout the present study; details of its architecture are deferred to the supplementary information (Sec.~\ref{sec:supp-oracle}). The frozen GNN model maps a valid quaternary composition $\xcomp \in \mathcal{X}$ to a predicted bulk modulus $\oracle(\xcomp)$. The inverse-design problem considered here is therefore
\begin{equation}
\xcomp^\star
=
\argmax_{\xcomp \in \mathcal{X}}
\oracle(\xcomp).
\label{eq:oracle-objective}
\end{equation}
Importantly, Eq.~\ref{eq:oracle-objective} defines optimization with respect to the frozen GNN-oracle rather than direct optimization of a DFT-calculated or experimentally measured bulk modulus. The reference dataset used to train the binary latent representation is treated as fixed offline information, and its property labels are obtained from the same GNN-oracle. Newly proposed compositions are likewise evaluated with this oracle. The oracle-call budgets introduced below therefore refer to additional unique oracle evaluations performed during the search, rather than to the offline reference data used to construct the latent representation. This distinction is particularly relevant when interpreting comparisons between methods that operate directly in composition space and methods that operate in the pretrained binary latent space.
All optimization results reported below should consequently be interpreted as optimization of a proxy objective. Because the oracle is a learned approximation to the physical property, a composition that is optimal, or near-optimal, under $\oracle(\xcomp)$ need not be optimal with respect to its DFT-calculated or experimentally measured bulk modulus. We therefore treat the present study as a methodological benchmark of inverse-design strategies under a fixed learned objective. The systematic error of the oracle, its behavior in the high-bulk-modulus tail, and the implications for subsequent DFT or experimental validation are discussed in the supplementary information (Sec.~\ref{sec:supp-oracle-bias}).

\section{QUBO-compatible inverse-design workflow}
\label{sec:workflow}

The goal of the workflow is to convert the frozen \lesets{} forward predictor
defined in Sec.~\ref{sec:problem-setting} into an inverse-design procedure that
remains compatible with binary quadratic optimization. The workflow combines a
binary variational autoencoder (\bvae{}), active learning in binary latent space,
a quadratic factorization-machine (FM) surrogate, and a final QUBO optimization
step.

\begin{figure}[t]
    \centering
    \includegraphics[width=\textwidth]{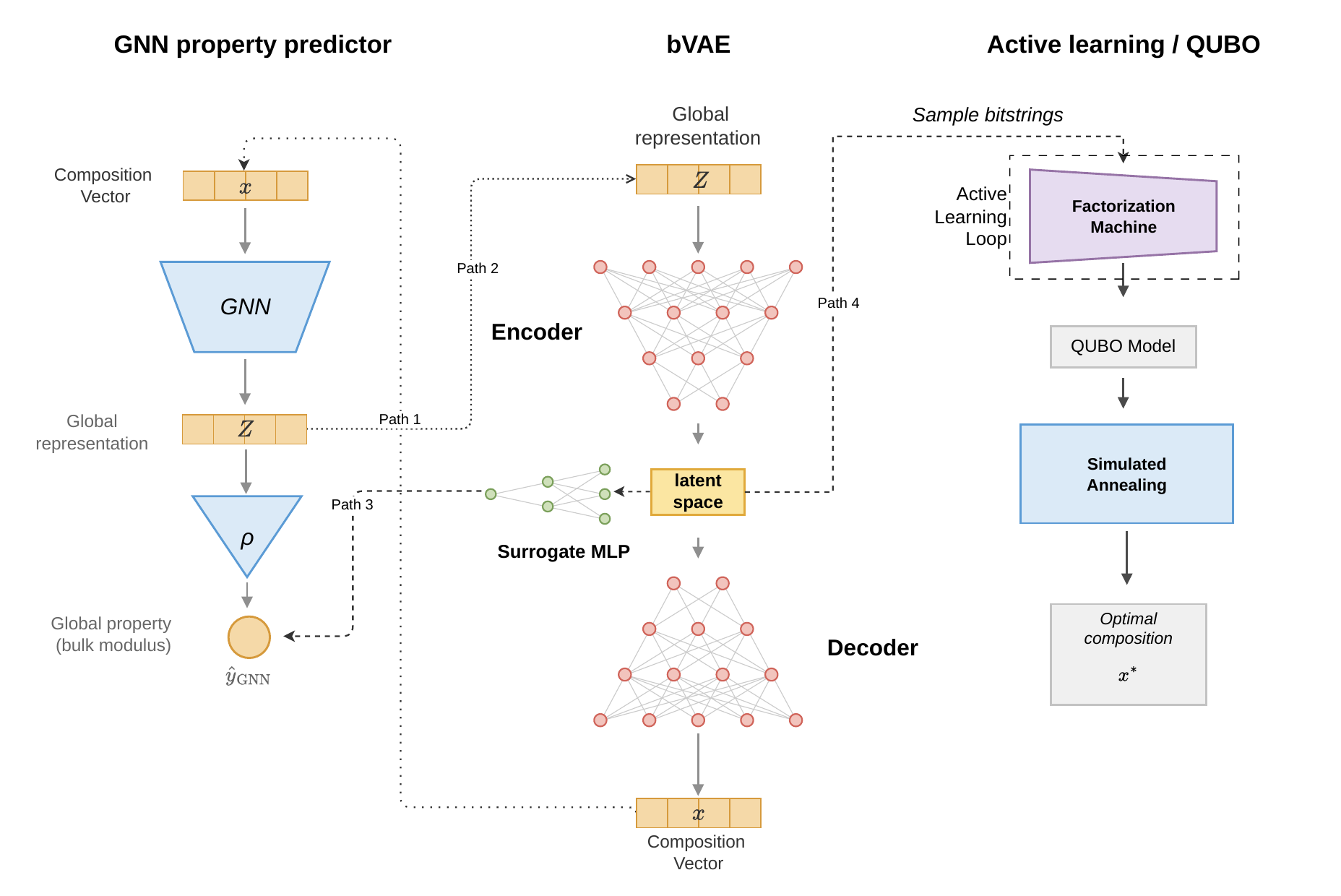}
    \caption{
       Overview of the QUBO-compatible inverse-design workflow, organized in three
    stages (columns). \emph{GNN property predictor:} the frozen \lesets{} model
    maps an alloy composition to a global representation $\bm{Z}$ and a predicted
    bulk modulus. \emph{bVAE:} the global representation $\bm{Z}$ is encoded into
    a binary latent code $\zbin \in \{0,1\}^{B}$ and decoded back to relaxed
    composition vectors, which are projected to valid quaternary compositions
    before oracle evaluation; an auxiliary surrogate MLP predicts properties from
    the latent code. \emph{Active learning / QUBO:} a BinaryFM ensemble is trained
    on evaluated latent-code/oracle-score pairs, and the final averaged FM is
    exported as a QUBO and solved by simulated annealing for verification. The
    numbered paths mark the key data flows: a decoded composition scored by the
    oracle (path~1), $\bm{Z}$ routed into the bVAE (path~2), the auxiliary
    surrogate predicting the bulk modulus (path~3), and latent codes sampled into
    the active-learning FM/QUBO stage (path~4).
    }
    \label{fig:workflow-schematic}
\end{figure}

A schematic overview is shown in Fig.~\ref{fig:workflow-schematic}. The overall
workflow has three main consecutive parts. First, a frozen \lesets{} maps an alloy composition to a global representation $\bm{Z}$, which represents the structure of the candidate molecule in terms of the chemical bonds, and a predicted bulk modulus.
Second, the frozen global representation $\bm{Z}$ is provided as input to a
\bvae{} (Fig.~\ref{fig:workflow-schematic}, path~2), whose encoder parametrizes a distribution $q_{\psi}(\bm{z}\mid\bm{Z})$ over binary latent codes $\bm{z} \in \{0,1\}^B$. A latent code is sampled from this distribution and
passed to the decoder, which maps it to a relaxed composition representation
over the full element vocabulary. This relaxed composition is subsequently
projected onto the valid quaternary composition domain before evaluation by the
\lesets{} (path~1). In parallel, an auxiliary multilayer-perceptron (MLP) surrogate is trained jointly with the \bvae{} to predict the oracle property from the latent representation (path~3). 
Third, during active learning an FM ensemble is trained on latent-code/oracle-score pairs sampled from the binary latent space (path~4). The ensemble is used to rank new candidate codes, while the final quadratic surrogate is exported as a QUBO and solved by simulated annealing to obtain QUBO-derived candidates for oracle verification.

\subsection{Oracle-derived representation data}
\label{sec:workflow-lesets-oracle}

The frozen \lesets{} model is used in two ways. Firstly, before optimization, it
provides the representation and label data used to train the binary latent
model. For each reference composition $\xcomp_n$, we store the triplet
\begin{equation}
    \left(
    \bm{Z}_n,
    \xcomp_n,
    y_n
    \right)
    =
    \left(
    \bm{Z}(\xcomp_n),
    \xcomp_n,
    \oracle(\xcomp_n)
    \right),
    \label{eq:lesets-triplets}
\end{equation}
where $\bm{Z}_n \in \mathbb{R}^{32}$ is the frozen \lesets{} global
representation, $\xcomp_n$ is the corresponding 15-dimensional composition
vector, and $y_n$ is the frozen-\lesets{} bulk-modulus prediction.

Secondly, during inverse design, the same GNN is queried as the oracle for newly decoded candidate compositions (Fig.~\ref{fig:workflow-schematic}, path~1). This ensures that all training labels used by the inverse-design workflow and all newly evaluated candidate scores are drawn from the same learned objective. As introduced in Sec.~\ref{sec:problem-setting}, the reference data constitutes fixed offline information, whereas newly evaluated candidates consume the online oracle-call budget used in the benchmark.

\subsection{Property-guided binary latent representation}
\label{sec:workflow-bvae}

Direct optimization in the continuous composition space is not naturally
compatible with QUBO solvers. We therefore introduce a binary latent
representation. Binary representation learning also provides a finite set of binary variables on which a quadratic surrogate can be defined directly.
Related work has shown that \bvae{}--FM workflows can provide competitive optimization performance relative to continuous-latent VAE approaches while enabling direct QUBO/Ising formulations~\citep{mao2023chemical,wilson2021machine}.

The \bvae{} encoder maps the frozen \lesets{} global representation
$\bm{Z}(\xcomp)$ (Fig.~\ref{fig:workflow-schematic}, path~2) of a composition vector $\xcomp$ to the parameters
of an approximate posterior distribution
$q_{\psi}(\zbin \mid \bm{Z}(\xcomp))$ over binary latent codes
$\zbin \in \{0,1\}^{B}$.

In the main experiments, $B=32$. The decoder maps a latent code to a relaxed 15-dimensional composition
representation over the fixed element vocabulary. Before oracle evaluation,
this relaxed output is projected to the valid quaternary composition space:
the four largest decoded elemental fractions are retained, all other fractions
are set to zero, and the retained fractions are renormalized to sum to one.

The \bvae{} is trained with a reconstruction loss, a binary latent regularization term, and an auxiliary property-prediction loss following surrogate-guided generative design~\citep{wang2020deep}. The auxiliary property surrogate (MLP) is used only during representation learning to organize the latent space along property-relevant directions. It is not used as the active-learning surrogate during inverse design. Consequently, the binary representation used in the subsequent optimization is property-informed before the online active-learning stage begins. Details of the \bvae{} architecture and training objective are given in Sec.~\ref{sec:supp-bvae-details}.

After training, the \bvae{} is frozen, following the general strategy used in related binary-latent inverse-design workflows~\citep{wilson2021machine,mao2023chemical}. Its encoder provides an aggregated posterior over binary latent codes for the reference data, and its decoder provides the map from binary latent codes to candidate alloy compositions.
Keeping the binary latent representation fixed is a
deliberate design choice: it defines a stable binary latent geometry, and hence
a stable QUBO variable set that is shared across all latent-space methods, so
that the benchmark compares search strategies rather than moving
representations. It also cleanly separates offline representation learning from the subsequent online optimization procedure.
~Retraining the encoder--decoder at every active-learning
iteration would also be computationally disproportionate relative to the
lightweight FM surrogate refit.

\subsection{Active learning with a quadratic factorization machine}
\label{sec:workflow-active-learning}

The active-learning loop is performed in binary latent space. At
iteration $t$, the workflow has accumulated a set of evaluated latent codes and
their oracle scores. An ensemble of $M$ quadratic factorization-machine
surrogates \cite{Rendle2010} is trained on the evaluated pairs. The ensemble is a bagged (bootstrap) ensemble rather than a cross-validation split: each of the $M$
surrogates is fit on an independent bootstrap resample drawn with replacement
(covering a fraction $\approx 0.9$ of the evaluated pairs), which provides an
empirical ensemble-based estimate of predictive variability.
For a binary latent code $\zbin \in \{0,1\}^{B}$, a single second-order FM predicts
\begin{equation}
    \hat{f}(\zbin)
    =
    w_0
    +
    \sum_{i=1}^{B} w_i z_i
    +
    \sum_{i<j}
    J_{ij} z_i z_j,
    \qquad
    J_{ij} = \langle \mathbf{v}_i,\mathbf{v}_j\rangle ,
    \label{eq:fm-explicit}
\end{equation}
where $w_0$ is a constant offset, $w_i$ are linear coefficients, and
$\mathbf{v}_i \in \mathbb{R}^{d_{\mathrm{FM}}}$ are learned vectors
associated with the latent variables. In the present workflow,
$d_{\mathrm{FM}}=8$. Instead of learning an independent coefficient for every
pair of latent variables, the FM parameterizes the pairwise interaction
coefficient $J_{ij}$ through the inner product of the corresponding
vectors. This factorized representation provides a compact model of pairwise
interactions. 

This model is distinct from the auxiliary property surrogate used during bVAE
training. The FM ensemble is trained only during inverse design, using oracle
evaluations collected by active learning. The first ensemble is trained on an
initialization set of oracle-evaluated latent codes sampled purely from the
broad hybrid distribution (Sec.~\ref{sec:workflow-candidate-generation}), with
no perturbation component; local perturbations are introduced only in
subsequent active-learning iterations, once high-performing codes are
available. In the main benchmark, this initialization phase uses $20\%$ of the
total oracle-call budget.
The ensemble predictions define a mean $\mu_t(\zbin)$ and standard deviation
$\sigma_t(\zbin)$ for each candidate latent code. Candidates are ranked by the
upper-confidence-bound (UCB) acquisition function
\begin{equation}
    a_t(\zbin)
    =
    \mu_t(\zbin)
    +
    \beta \sigma_t(\zbin),
    \label{eq:ucb-acquisition}
\end{equation}
where $\beta$ controls the exploration--exploitation trade-off.

\subsection{Candidate-pool construction and local perturbations}
\label{sec:workflow-candidate-generation}

The acquisition function is evaluated on a finite candidate pool
$\mathcal{C}_t \subset \{0,1\}^{B}$. The construction of this pool is a central
design choice in the workflow. The pool contains a broad component and a perturbed component.

The broad component of the pool samples binary latent codes from a hybrid
distribution combining two sources. The first is the uniform Bernoulli prior on
$\{0,1\}^{B}$, i.e.\ each bit is an independent fair coin flip; this samples the
latent space uniformly, over both regions already populated by the reference
data and regions not yet visited. The second is the aggregated encoder
posterior: the per-bit Bernoulli activation probabilities produced by the
frozen \bvae{} encoder on the reference compositions, averaged over that
dataset into a single per-bit probability vector. Sampling from this distribution produces samples resembling the offline reference data.
~The mixing parameter $\alpha_{\mathrm{agg}}$ controls the
fraction of samples drawn from the aggregated posterior. In the final benchmark
configuration, $\alpha_{\mathrm{agg}} = 0.5$, so the broad component is an even
mix of uniform Bernoulli samples and aggregated-posterior samples.

In a second step, additional candidates are generated by locally perturbing the
best latent codes already evaluated by the oracle, providing local exploitation
capability. These perturbations are applied to the previously evaluated and
oracle-scored codes, not to the freshly drawn broad samples. At each iteration,
all previously evaluated codes are therefore ranked by their oracle scores. The
top $K$ codes are selected, and for each of them multiple perturbed copies are
generated by flipping a small number of randomly selected bits. In the main
benchmark, we use the top $K=16$ evaluated codes, generate $C=64$ perturbed
copies per top code, and flip between one and $R=3$ randomly selected bits per
copy. The perturbed copies are not explicitly deduplicated; identical or
previously seen codes that recur are collapsed at evaluation time by the oracle
cache (Sec.~\ref{sec:benchmark-methods}), so they do not consume additional
oracle-call budget.

The full candidate pool is the union of the broad and perturbed components. The
pool determines which candidates are made available to the
acquisition function; the final selection remains model-based through the
FM-UCB score in Eq.~\eqref{eq:ucb-acquisition}. This design allows the workflow
to combine broad exploration of the binary latent space with local exploitation
around high-performing regions already discovered by the oracle.

At each active-learning iteration, the acquisition function selects a query
batch targeting a fixed number of new unique oracle evaluations (500 in the
main benchmark), so that the $20\%$-initialization
budget respects all (scheduled) iterations.

Algorithm~\ref{alg:active-learning-perturbation} summarizes the active-learning
loop used in the reference workflow. The key modification relative to a
prior-only candidate pool is the addition of local perturbations of the best
evaluated latent codes before FM-UCB ranking. This differs from previous
factorization-machine QUBO workflows~\citep{wilson2021machine,mao2023chemical},
in which the QUBO/Ising surrogate is re-solved at every iteration to propose the
next candidate. Here, candidate proposal during active learning is driven by
FM-UCB ranking over the hybrid broad-plus-perturbation pool, and QUBO solving is
reserved for the final verification step (Sec.~\ref{sec:workflow-qubo}).
This separation between acquisition and QUBO solving is deliberate: the active-learning stage uses ensemble uncertainty to guide data acquisition, whereas the final QUBO represents the mean quadratic surrogate without an uncertainty term.
The low-level oracle-cache and
budget-accounting details are omitted here and described in
Sec.~\ref{sec:benchmark-methods}.
After the final active-learning iteration, the last FM ensemble is averaged and
exported as the QUBO objective described in Sec.~\ref{sec:workflow-qubo}.

\begin{algorithm}[t]
\caption{Active learning with local latent perturbations}
\label{alg:active-learning-perturbation}
\begin{algorithmic}[1]

\REQUIRE Initial evaluated set $\mathcal{D}_0$, frozen oracle $f$, decoder $D$,
number of iterations $T$, batch size $b$, perturbation parameters $K,C,R$,
FM ensemble size $M$, UCB parameter $\beta$

\STATE $\mathcal{D} \leftarrow \mathcal{D}_0$

\FOR{$t = 1,\ldots,T$}
    \STATE Train FM ensemble $\{\hat f_m\}_{m=1}^{M}$ on $\mathcal{D}$
    \STATE Sample broad latent pool $\mathcal{C}_{\mathrm{broad}}$
    \STATE Select top-$K$ evaluated codes from $\mathcal{D}$ (by oracle score)
    \STATE Generate perturbed pool $\mathcal{C}_{\mathrm{pert}}$ with $C$ copies of each top code, each flipping $1,\ldots,R$ bits
    \STATE $\mathcal{C} \leftarrow \mathcal{C}_{\mathrm{broad}} \cup \mathcal{C}_{\mathrm{pert}}$
    \STATE Compute ensemble mean $\mu(z)$ and uncertainty $\sigma(z)$ for all $z \in \mathcal{C}$
    \STATE Select query batch
    \[
        \mathcal{Q}
        =
        \operatorname{TopB}_{z \in \mathcal{C}}
        \left[
        \mu(z) + \beta \sigma(z)
        \right]
    \]
    \STATE Decode and project each $z \in \mathcal{Q}$ to a valid composition $x$
    \STATE Evaluate $y = f(x)$ for each query and update
    \[
        \mathcal{D} \leftarrow \mathcal{D} \cup \{(z,y): z \in \mathcal{Q}\}
    \]
\ENDFOR

\RETURN $\mathcal{D}$

\end{algorithmic}
\end{algorithm}

\subsection{QUBO optimization of the learned surrogate}
\label{sec:workflow-qubo}

After active learning, the final FM ensemble is converted into a
single quadratic surrogate by averaging the corresponding offset, linear coefficients, and pairwise quadratic coefficients across ensemble members. Let
$\bar{w}_0$, $\bar{w}_i$, and $\bar{J}_{ij}$ denote the averaged offset, linear
coefficients, and pairwise coefficients, respectively. The resulting averaged surrogate is
quadratic in the binary latent variables and is therefore directly
QUBO-compatible.

Maximizing the averaged FM surrogate is equivalent to minimizing its negative.
The QUBO problem is therefore to find the binary latent code that minimizes the
negative averaged surrogate,
\begin{equation}
    \zbin^\star
    =
    \argmin_{\zbin \in \{0,1\}^{B}}
    \left(
    -
    \sum_i \bar{w}_i z_i
    -
    \sum_{i<j}
    \bar{J}_{ij} z_i z_j
    \right),
    \label{eq:qubo-problem}
\end{equation}
where the constant offset $\bar{w}_0$ has been dropped because it does not affect
the minimizer. This objective is quadratic in the binary variables and is
directly compatible with QUBO/Ising solvers.

In this work, the QUBO is solved using simulated annealing, which returns a
sample set of binary codes. The best samples (up to five in the main benchmark)
are decoded, projected to valid quaternary compositions, and verified with the
frozen \lesets{}. Because simulated annealing is a heuristic solver, the returned sample is not assumed to constitute a certified global optimum of Eq.~\eqref{eq:qubo-problem}.
~Importantly, although the final design selections are
made on the basis of the surrogate prediction, the bulk moduli of the selected
designs are determined with the original \lesets{}. This is crucial for
benchmarking different surrogate models and search strategies on a common,
consistent objective.
In the final benchmark configuration, the QUBO optimization step is kept small.
Most high-scoring candidates are found during active learning, while QUBO
optimization demonstrates that the learned FM surrogate can be exported to a
QUBO-compatible objective and solved by a compatible backend.

It is worth clarifying the distinct roles of the active-learning acquisition in
Eq.~\eqref{eq:ucb-acquisition} and the QUBO in Eq.~\eqref{eq:qubo-problem}, since
the two optimize different objectives. The acquisition function ranks a
\emph{finite, sampled} candidate pool by the upper-confidence bound
$\mu_t + \beta\sigma_t$, deliberately trading exploitation of the surrogate mean
against exploration of its uncertainty; it does not directly optimize the mean surrogate over the full binary domain, and its reach is limited to the broad and perturbed codes present in
the pool at iteration $t$. The QUBO, in contrast, is defined by the negative mean of the final averaged FM surrogate over the entire binary hypercube $\{0,1\}^{B}$, with no uncertainty term and no restriction to the sampled candidate pool. The simulated-annealing solver can therefore explore solutions outside the candidate pools considered during active learning, although it does not guarantee global optimality.
~Solving the QUBO is therefore not redundant with active learning: it provides a distinct optimization endpoint for the learned surrogate and realizes the method's structural motivation, namely that the final surrogate can be exported directly to QUBO/Ising form and solved on a compatible backend.

This also explains why the best QUBO-derived candidate does not simply supersede the active-learning result. The averaged surrogate is trained on oracle evaluations concentrated in the region that active learning has populated, so it is most strongly supported by training data in those regions and may extrapolate in sparsely sampled parts of the latent space.
A low QUBO energy obtained in such a region can therefore reflect surrogate extrapolation rather than a genuinely high oracle score.
Once decoded, projected, and verified with the \lesets{}, these candidates do not, in our experiments, exceed the best designs already found during active learning. Consistent with this, solving the QUBO at every active-learning iteration rather than once at the end does not improve the final result (Sec.~\ref{sec:results-workflow-variants}). The final QUBO is thus retained as a full-domain surrogate-optimization and structural-compatibility endpoint that verifies the exported surrogate, rather than as the primary source of the best-scoring designs.
\section{Benchmark design and methods}
\label{sec:benchmark-methods}

We benchmark the proposed QUBO-compatible workflow in two complementary settings. The primary controlled comparison is between search methods that operate in the same pretrained binary latent representation. In addition, we include methods that operate directly in composition space as contextual baselines for the overall optimization performance of the binary-latent workflow. All online search methods use the same frozen GNN-model \textsc{LESets} \cite{zhang2025graph} as the property oracle and are compared under matched budgets of additional unique oracle evaluations. As discussed in Sec.~\ref{sec:problem-setting}, however, the latent-space methods inherit the property-guided representation learned from the fixed offline reference dataset, whereas the composition-space methods do not. Comparisons between the two search spaces should therefore not be interpreted as controlled comparisons of the search space alone. A repeated proposal that decodes to a composition already evaluated within the same run is served from a cache and is not counted as a new oracle call. This accounting is particularly important for latent-space methods, where different binary codes can decode to the same projected quaternary composition.

Accordingly, the benchmark addresses two questions. First, within the shared binary latent space, we examine the contributions of model-based FM-UCB selection and local latent perturbations to search performance. Second, we compare the resulting workflow with direct composition-space optimization methods under the same additional oracle-evaluation budget to place its overall performance in context. The methods considered in the benchmark are summarized in Table~\ref{tab:benchmark-methods}. Composition-space methods operate directly on valid quaternary alloy compositions, whereas all latent-space methods use the same pretrained bVAE decoder to map binary latent codes to valid compositions.

\begin{table}[ht]
    \centering
    \caption{
    Benchmark methods used to evaluate the QUBO-compatible inverse-design workflow.
    All methods are compared under matched unique-oracle-call budgets, where one oracle call corresponds to one previously unevaluated candidate composition evaluated by the frozen \lesets{} predictor.
    }
    \label{tab:benchmark-methods}
    \small
    \begin{tabular}{p{0.12\linewidth} p{0.15\linewidth} p{0.25\linewidth} p{0.2\linewidth} p{0.12\linewidth}}
        \toprule
        \textbf{Method} &
        \textbf{Search space} &
        \textbf{Candidate generation} &
        \textbf{Selection strategy} &
        \textbf{QUBO-compatible?} \\
        \midrule

        \randomcomp{} &
        Composition space &
        Random valid quaternary compositions sampled from the allowed element set and composition simplex. &
        Uniform random selection. &
        No \\

        \gacomp{} &
        Composition space &
        Population of valid quaternary compositions evolved by crossover, mutation, and repair to enforce constraints. &
        Genetic selection based on \lesets{} score. &
        No \\

        \rfucbcomp{} &
        Composition space &
        Fresh random pool of valid quaternary compositions generated at each iteration. &
        Random-forest UCB acquisition,
        $\mu(\xcomp) + \beta \sigma(\xcomp)$. &
        No \\

        \randomlatent{} &
        Binary latent space &
        Binary codes sampled from a hybrid distribution combining the Bernoulli prior and aggregated \bvae{} encoder posterior. &
        Uniform random selection after decoding and projection to valid compositions. &
        No \\

        \randompertlatent{} &
        Binary latent space &
        Mixture of random latent samples and local perturbations of the best previously evaluated latent codes. &
        Uniform random selection from the random/perturbed candidate mixture. &
        No \\

        \galatent{} &
        Binary latent space &
        Population of binary latent codes evolved by uniform crossover and bit-flip mutation, followed by decoding. &
        Genetic selection based on decoded \lesets{} score. &
        No \\

        \workflownopert{} &
        Binary latent space &
        Active-learning candidate pool sampled from the hybrid Bernoulli-prior / aggregated-posterior distribution. &
        Ensemble \binaryfm{} UCB acquisition,
        $\mu(\zbin) + \beta \sigma(\zbin)$. &
        Yes \\

        \workflowpert{} &
        Binary latent space &
        Active-learning candidate pool combining broad hybrid latent samples with local perturbations of top evaluated codes. &
        Ensemble \binaryfm{} UCB acquisition followed by final QUBO verification. &
        Yes \\

        \bottomrule
    \end{tabular}
\end{table}

\subsection{Reference workflow and ablation study}

The full workflow is the QUBO-compatible active-learning method described in Sec.~\ref{sec:workflow}. It uses the pretrained bVAE decoder, an ensemble of FM surrogates, FM-UCB acquisition, the hybrid broad-plus-perturbative candidate pool, and final QUBO optimization of the learned surrogate. We denote this method Workflow. 

To assess the role of the perturbative candidate-pool component, we also evaluate Workflow(no pert). This method is identical to Workflow except that the perturbative component is removed from the active-learning candidate pool. The acquisition function remains model-based, and candidates are still ranked using the FM-UCB score, but the pool contains only broad latent samples from the prior/aggregated-posterior mixture.

Together with Random(latent) and Random+Pert(latent), these workflow variants probe the contributions of broad latent sampling, local perturbations, and model-based selection. Comparing Workflow(no pert) with Workflow directly isolates the effect of adding local perturbations to the FM-UCB candidate pool. Comparing Random(latent) with Random+Pert(latent) provides a model-free assessment of the benefit of locally exploring neighborhoods of previously high-scoring latent codes. Finally, comparing Random+Pert(latent) with Workflow indicates the additional benefit obtained when perturbation-enhanced candidate generation is combined with FM-UCB ranking. This last comparison is not a strict one-factor ablation, because the two methods differ in their initialization and candidate-selection protocol, and is therefore interpreted as a complementary assessment rather than an exact isolation of the FM-UCB contribution.

\subsection{Latent-space search methods}
\label{sec:benchmark-latent}

The latent-space methods perform a search over binary latent codes
$\zbin \in \{0,1\}^{B}$ using the pretrained bVAE decoder. Each proposed
latent code is decoded and projected to a valid composition using the projection
described in Sec.~\ref{sec:workflow-bvae}, and the resulting composition is
evaluated with the frozen \lesets{}. Because these methods share the same property-guided binary representation, their comparison provides the main controlled benchmark of alternative search strategies in latent space.

\paragraph{Random search.}
The Random(latent) method samples binary latent codes from the same broad latent
distribution used by the workflow candidate generator, namely a mixture of the
uniform Bernoulli prior and the aggregated bVAE encoder posterior. Each sampled
code is decoded, projected to $\mathcal{X}$, and evaluated with the frozen
\lesets{}. No surrogate model, acquisition function, evolutionary update,
or QUBO solve is used. This provides the reference performance obtained from the pretrained latent representation without adaptive search.

\paragraph{Genetic algorithm.}
The GA(latent) method applies a genetic algorithm directly to binary latent
codes. The population is initialized by sampling from the broad latent
distribution. Fitness is computed by decoding each latent code to a valid
composition and evaluating the \lesets{}.
At each generation, parents are selected by tournament selection. Offspring are
generated using uniform crossover followed by bit-flip mutation. The mutation
operator flips each bit independently with a fixed mutation probability and
ensures that at least one bit is flipped. Since the representation is binary,
offspring are valid latent-space candidates by construction. After evaluation,
the next population is formed by keeping the highest-scoring candidates from
the union of the parent population and the offspring. This method tests the
strength of a classical black-box optimizer in the same binary latent space
used by the proposed workflow.

\paragraph{Random search with local perturbations.}
Random+Pert(latent) is included to assess the effect of local perturbations
without model-based candidate ranking. The method uses the same type
of perturbations as the proposed workflow, but removes the FM ensemble, UCB
acquisition function, and final QUBO optimization stage.
The method begins with a random latent warm-start phase, using approximately
15\% of the evaluation budget. After initialization, each iteration evaluates a
batch consisting of 80\% broad latent samples and 20\% perturbations of the
best latent codes observed so far. Previously scored latent codes are ranked by
their oracle values, the top codes are selected, and perturbed copies are
produced by flipping a small number of randomly selected bits. The resulting
candidates are decoded, projected, and evaluated directly with the
\lesets{}.
This method tests whether the local perturbation mechanism alone can account for a substantial part of the improvement observed in the full workflow. Unlike the proposed workflow,
Random+Pert(latent) does not train a surrogate and does not rank candidates by
an acquisition function. Perturbed candidates are evaluated solely because they
lie near previously high-scoring latent codes.

\subsection{Composition-space search methods}

The composition-space methods operate directly on the constrained composition space $\mathcal{X}$ defined in Eq.~\eqref{eq:composition-space}. Each candidate is represented as a 15-dimensional composition vector $\xcomp \in \mathbb{R}^{15}$ over the fixed element pool, with exactly four nonzero entries, non-negative fractions, and total fraction equal to one. These methods do not use the pretrained binary representation and are included to place the overall optimization performance of the QUBO-compatible workflow in the context of direct search over the original composition domain.

\paragraph{Random search.}
The Random(comp) method samples valid quaternary compositions independently. For
each proposal, four distinct elements are selected uniformly from the
15-element pool and their fractions are sampled from a Dirichlet distribution
with unit concentration parameters. The resulting fractions are embedded into a
15-dimensional composition vector and evaluated using the \lesets{}. This method does not use a surrogate model, latent representation,
evolutionary update, or local search.

\paragraph{Genetic algorithm.}
The composition-space genetic algorithm, denoted GA(comp), is an evolutionary
search method over valid composition vectors \citep{shapiro1999genetic}. The
population is initialized with random valid quaternary compositions. At each
generation, parents are selected by tournament selection using their
\lesets{} values as fitness.
Offspring are produced by convex crossover between two parent compositions,
followed by mutation. The crossover coefficient is sampled from an intermediate
range, so the child is a blend of the two parent compositions. Mutation can
either swap one active element for an inactive element or perturb the active
elemental fractions by multiplicative noise. After crossover and mutation, a
repair step projects the child back to a valid quaternary composition by
retaining exactly four active elements, clipping negative entries, and
renormalizing the active fractions to sum to one. The next population is formed
by keeping the highest-scoring candidates from the union of the current
population and the newly evaluated offspring.

\paragraph{Random-Forest-UCB search.}
RF-UCB(comp) uses a Random Forest (RF) surrogate with upper-confidence-bound
acquisition to guide composition-space search \citep{breiman2001random}. The
method begins from an initial set of randomly sampled valid compositions. At
iteration $t$, the Random Forest is trained on all observed
composition--oracle pairs. A fresh pool of valid quaternary compositions is then
generated, excluding compositions that have already been selected for
evaluation.
For each candidate $\xcomp$ in this fresh pool, the predictive mean is computed
as the average prediction over trees, and the tree-to-tree predictive variability is quantified as the standard deviation of the tree predictions. Candidates are ranked using the same
UCB form as in Eq.~\eqref{eq:ucb-acquisition}, but applied in composition space
with the Random Forest mean and uncertainty. The highest-acquisition candidates
are evaluated with the \lesets{}. The selected candidates are
added to the observed dataset, and the forest is refit at the next iteration.
The surrogate is therefore cumulative: at every iteration the Random Forest is
retrained on the full history of composition--oracle pairs observed so far, so
progress made in earlier iterations is retained. What is regenerated at each
iteration is only the candidate pool: rather than ranking a fixed, precomputed
library of compositions, the method draws a fresh set of random valid
quaternary compositions (excluding those already evaluated) and ranks these with
the current surrogate. This keeps the accessible search space effectively
unbounded across iterations while still guiding evaluation with all accumulated
data.

\subsection{Initialization.}
The methods differ in how they are seeded, and in all cases the initialization consumes part of the same online unique-oracle-call budget. The two random baselines, Random(comp) and Random(latent), have no separate initialization phase and simply draw independent proposals from their respective distributions. The genetic algorithms are seeded with an initial population of $128$ candidates: random valid quaternary compositions for GA(comp), and codes sampled from the broad latent distribution for GA(latent). RF-UCB(comp) begins from $256$ random valid quaternary compositions before its first surrogate fit. Random+Pert(latent) uses a random latent warm-start of approximately $15\%$ of the budget (at least $50$ proposals), and the QUBO-compatible workflow uses an initialization phase of $20\%$ of the budget drawn from the broad latent distribution (Sec.~\ref{sec:workflow-active-learning}). Counting all initialization evaluations against the same online budget ensures that no method receives uncounted oracle evaluations during the search. It does not imply equal prior information across the two representation classes: the latent-space methods inherit the fixed offline-trained bVAE, whereas the composition-space methods operate without this pretrained representation. Cross-space comparisons are therefore used to contextualize overall performance, while mechanistic conclusions about the search strategy are drawn primarily from comparisons within the shared latent space.

\section{Results}
\label{sec:results}

We evaluate the QUBO-compatible workflow on the bulk-modulus inverse-design task using the frozen GNN-predictor as the common oracle. Performance is measured as the best score found up to a given budget of additional unique oracle evaluations, as described in Sec.~\ref{sec:benchmark-budget-cache}.
Following the benchmark design of Sec.~\ref{sec:benchmark-methods}, we distinguish
between two types of comparisons. The primary controlled benchmark compares search
strategies operating in the same pretrained binary latent representation.
Composition-space methods are reported separately as contextual baselines for the
overall optimization performance of the binary-latent workflow.

\subsection{Search performance in binary latent space}
\label{sec:results-benchmark-performance}
\begin{figure}[ht]
    \centering
    \includegraphics[width=\textwidth]{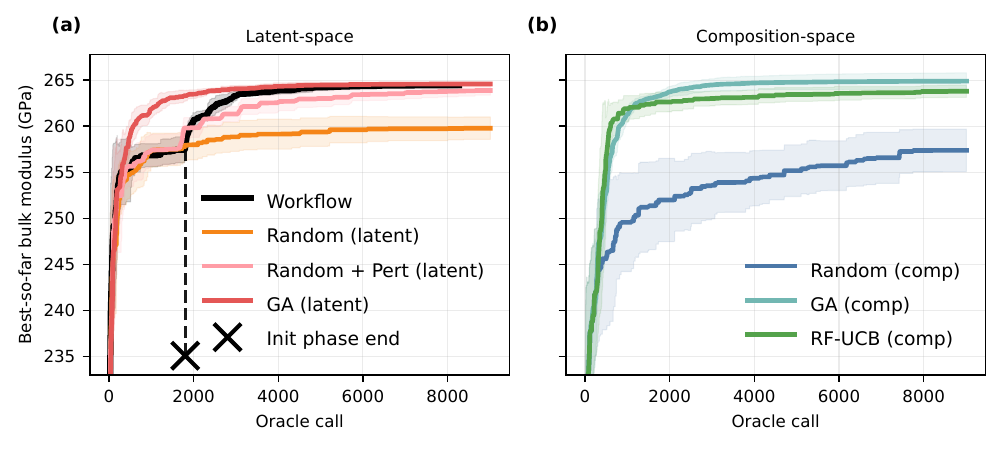}
    \caption{
   Best-so-far \lesets{} bulk score as a function of unique oracle evaluations.
    (a) Workflow compared with latent-space searches. (b) Composition-space
    baselines. Curves show the mean across 20 seeds
    and shaded regions the sample standard deviation across those seeds, computed pointwise at each evaluation.
    }
    \label{fig:results-best-so-far}
\end{figure}

We first compare methods that operate in the same pretrained binary latent space.
Figure~\ref{fig:results-best-so-far}a shows the best-so-far oracle score as a
function of the number of unique oracle evaluations. Random(latent) establishes
the performance obtained from the learned representation without adaptive
search, whereas Random+Pert(latent), GA(latent), Workflow(no pert), and the full
Workflow introduce progressively different mechanisms for exploiting the binary
search space.

All adaptive latent-space methods improve rapidly during the early part of the
search. Random+Pert(latent) substantially exceeds purely random latent sampling,
showing that local exploration around previously high-scoring codes is already
effective without a learned acquisition model. GA(latent) and the full workflow
reach the highest score regime among the latent-space methods.

\subsection{Role of local perturbations and model-based selection}
\label{sec:results-perturbation-ablation}
The clearest workflow-specific effect is associated with the construction of the
active-learning candidate pool described in Sec.~\ref{sec:workflow-candidate-generation}. 
In the reference workflow, broad latent samples
are supplemented by local bit-flip perturbations of high-performing previously
evaluated codes. Workflow(no pert) provides a direct one-factor ablation of this
mechanism: both variants use the same pretrained bVAE representation, FM
ensemble, UCB acquisition function, and final QUBO optimization, and differ only
in whether local perturbations are added to the candidate pool.

\begin{figure}[ht]
    \centering
    \includegraphics[width=\textwidth]{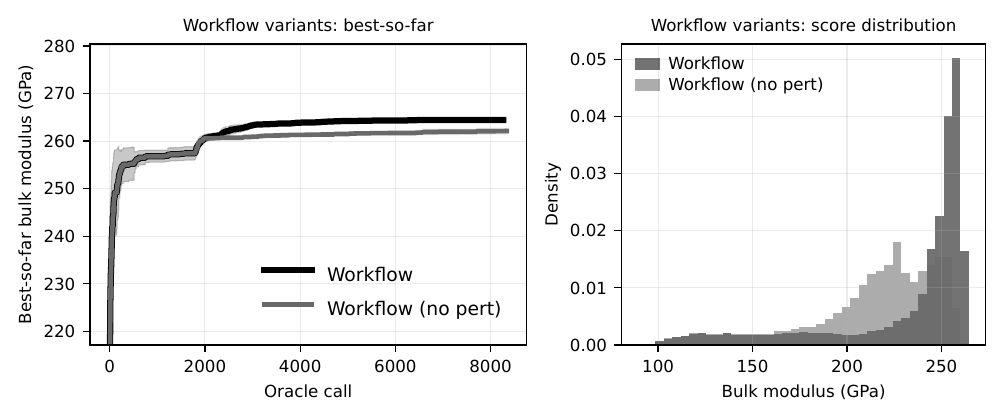}
    \caption{
    Effect of local latent perturbations on the QUBO-compatible workflow.
    Left: best-so-far oracle score for Workflow and
    Workflow(no pert). Right: distributions of the oracle scores of evaluated
    candidates. Adding local perturbations shifts the evaluated population
    toward higher \lesets{} scores and increases the best score reached under the
    same oracle-call budget.
    }
    \label{fig:results-perturbation-ablation}
\end{figure}

Figure~\ref{fig:results-perturbation-ablation} shows that adding perturbations
shifts oracle evaluations strongly toward the high-score region and increases
the final best score. Without perturbations, the active-learning loop evaluates
a broader population of moderate-score candidates and reaches a lower plateau.
The acquisition function can only rank candidates that are present in its
finite candidate pool; broad sampling alone therefore does not expose enough
high-quality local neighborhoods for FM-UCB selection. Local perturbations
provide a simple exploitation mechanism that enriches the pool around promising
regions before model-based ranking is applied.

The Random+Pert(latent) baseline provides complementary evidence for this
interpretation. Its strong performance shows that local perturbation is useful
even without the FM ensemble or UCB acquisition. The full workflow nevertheless
reaches a higher score, consistent with an additional benefit from
model-based ranking within the enriched candidate pool. Because
Random+Pert(latent) and the full workflow also differ in initialization and
candidate-selection protocol, this comparison should not be interpreted as a
strict one-factor isolation of the FM-UCB contribution.

\begin{figure}[ht]
    \centering
    \includegraphics[width=\textwidth]{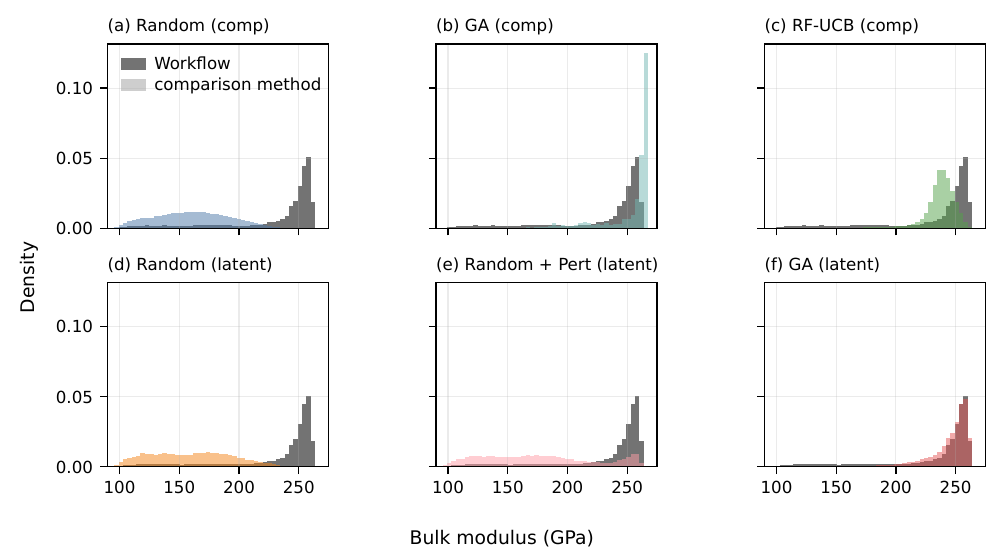}
    \caption{Pairwise score distributions comparing Workflow with each
    baseline. Each panel shows the density of \lesets{} scores for evaluated
    candidates: (a) Random(comp), (b) GA(comp), (c) RF-UCB(comp),
    (d) Random(latent), (e) Random+Pert(latent), and (f) GA(latent).
    }
    \label{fig:results-score-distributions}
\end{figure}

A broader view of the evaluated-score distributions is shown in
Fig.~\ref{fig:results-score-distributions}. Random search covers a wide range of
oracle scores, whereas adaptive and evolutionary methods increasingly
concentrate their evaluations near the high-score region. Within the shared
latent space, the distributions reinforce the mechanism identified above:
Random+Pert(latent), GA(latent), and the perturbation-enhanced workflow devote a
larger fraction of their evaluations to high-scoring candidates than
Random(latent) or Workflow(no pert). The workflow nevertheless retains a broader
support than a purely exploitative local search because broad candidate sampling
is maintained throughout active learning.

\subsection{Comparison with composition-space optimization}
\label{sec:results-composition-context}
We next place the performance of the binary-latent workflow in the context of
direct optimization in the original composition space. As discussed in
Sec.~\ref{sec:benchmark-methods}, this is not a controlled comparison of the
search representation alone: the latent-space methods inherit the
property-guided bVAE trained from the fixed offline reference data, whereas the
composition-space methods do not. The comparison instead asks whether the
QUBO-compatible binary-latent workflow remains competitive with direct classical
optimization of the alloy-composition domain.

Random(latent) performs much better than Random(comp) showing that the property-guided binary representation supports effective optimization.
The composition-space genetic algorithm, GA(comp), achieves the highest mean
final oracle score in the present benchmark. RF-UCB(comp) also improves
substantially over random composition search, while Random(comp) is the weakest
overall baseline. The full binary-latent workflow approaches the performance of
GA(comp) and reaches a score regime similar to GA(latent). Thus, the present
results do not indicate an optimization advantage from imposing a binary
QUBO-compatible representation. Rather, they show that competitive inverse-design
performance can be retained while constraining the learned representation and
surrogate to a form that permits direct QUBO export.

Table~\ref{tab:results-final-scores} summarizes these endpoint results. The
latent-space methods provide the controlled comparison of search strategies,
while the composition-space methods provide the contextual reference for direct
classical optimization.

\begin{table}[ht]
    \centering
    \caption{
    Final best \lesets{}-predicted bulk modulus achieved by each method at the
    matched oracle-call budget, in GPa (mean $\pm$ s.d. across 20 independent
    random seeds). The latent-space methods provide the controlled comparison
    within the shared pretrained binary representation, while the
    composition-space methods serve as contextual baselines. GA(comp) achieves
    the highest mean overall, whereas the proposed Workflow reaches a similar
    regime to GA(latent) and outperforms Random(latent) and Workflow(no pert).
    The strong result for Random+Pert(latent) shows that local perturbations
    account for a substantial part of the improvement in the binary latent space.
    }
    \label{tab:results-final-scores}
    \begin{tabular}{lc}
        \hline
        Method & Final best (GPa) \\
        \hline
        \multicolumn{2}{l}{\emph{Latent space}} \\
        Workflow             & $264.46 \pm 0.13$ \\
        Workflow (no pert.)  & $262.23 \pm 0.68$ \\
        GA (latent)          & $264.57 \pm 0.19$ \\
        Random+Pert (latent) & $263.87 \pm 0.50$ \\
        Random (latent)      & $259.77 \pm 1.19$ \\
        \hline
        \multicolumn{2}{l}{\emph{Composition space}} \\
        GA (comp.)           & $264.88 \pm 0.87$ \\
        RF-UCB (comp.)       & $263.77 \pm 0.64$ \\
        Random (comp.)       & $257.36 \pm 2.26$ \\
        \hline
    \end{tabular}
\end{table}

\subsection{QUBO optimization of the learned surrogate}
\label{sec:results-qubo-endpoint}
The final stage of the workflow exports the averaged FM surrogate to the QUBO
defined in Sec.~\ref{sec:workflow-qubo} and solves the resulting binary
quadratic objective by simulated annealing. The returned binary codes are
decoded and projected to valid quaternary compositions before their bulk-modulus
scores are evaluated with the frozen \lesets{}. The QUBO stage therefore
tests whether the surrogate, learned during active learning, can be transferred
directly to a binary quadratic optimization backend and whether the resulting
surrogate-selected candidates retain high oracle scores.

In the present benchmark, the best QUBO-derived candidates generally do not
surpass the best candidates already discovered during active learning. Most of
the improvement in the oracle objective is obtained before the final QUBO solve,
particularly once local perturbations are included in the candidate pool. The
QUBO result should therefore be interpreted as demonstrating optimization of the
final learned quadratic surrogate over the binary domain, rather than as the
principal source of the best oracle-verified candidates. In particular, these
experiments do not establish a performance advantage from annealing or from
quantum optimization.

\subsection{Robustness to additional workflow variants}
\label{sec:results-workflow-variants}
We finally test whether the reference workflow depends sensitively on two
additional design choices: the fixed mixture of broad prior and
aggregated-posterior sampling, and the decision to solve the QUBO only after the
active-learning loop. Each variant is compared with its corresponding reference
workflow under the same frozen oracle, matched unique-oracle-call budget, and
the same 20 random seeds (Table~\ref{tab:workflow-variants}).

\begin{table}[ht]
    \centering
    \caption{
    Workflow variant experiments. Final best \lesets{}-predicted bulk modulus
    (GPa, mean $\pm$ s.d. across 20 seeds) at the matched oracle-call budget.
    Both variants are compared against the same reference-workflow run: it is
    configuration-identical in the two experiments, so a single set of 20 seeded
    runs serves as the common baseline. Both differences are well below the
    seed-to-seed spread ($0.13$--$0.29$~GPa) and far below the oracle's own
    prediction error ($\mathrm{RMSE} \approx 12$~GPa), so neither variant
    meaningfully changes the result.
    }
    \label{tab:workflow-variants}
    \begin{tabular}{llrr}
        \hline
        Experiment & Arm & Final best (GPa) & $\Delta$ \\
        \hline
        Mixing-rate schedule & Reference (fixed rate) & $264.460 \pm 0.129$ & --- \\
                             & Scheduled              & $264.360 \pm 0.290$ & $-0.100$ \\
        \hline
        Per-iteration QUBO   & Reference (end-only)   & $264.460 \pm 0.129$ & --- \\
                             & Every iteration        & $264.454 \pm 0.150$ & $-0.006$ \\
        \hline
    \end{tabular}
\end{table}

\paragraph{Scheduled exploration--exploitation mixing.}
The broad candidate distribution described in
Sec.~\ref{sec:workflow-candidate-generation} combines samples from a uniform Bernoulli prior
and the aggregated \bvae{} encoder posterior according to the mixing parameter
$\alpha_{\mathrm{agg}}$. The reference workflow uses a fixed value. To test
whether progressively shifting from broad exploration toward the
reference-data-informed posterior improves the search, we replace the constant
value by
\begin{equation}
\alpha_{\mathrm{agg}}(t)
=
\varepsilon
+
(1-2\varepsilon)
\left(1-e^{-t/\tau}\right),
\label{eq:agg-frac-schedule}
\end{equation}
with iteration index $t$, ramp constant $\tau=4$, and
$\varepsilon=0.05$.

Across 20 seeds, the scheduled variant reaches a final best score of
$264.4\pm0.3$~GPa, compared with $264.5\pm0.1$~GPa for its fixed-rate reference
workflow. The best-so-far trajectories largely overlap, and the difference in
the final means is small relative to the observed seed-to-seed variation. 
The scheduled variant is not slower to improve in the early iterations; 
the trajectories separate only in the later part of the search, 
where the fixed-rate reference retains a small lead. 
We therefore retain the simpler fixed mixing ratio for the reference workflow.

\paragraph{Per-iteration QUBO optimization.}
The reference workflow solves the averaged-FM QUBO only after active learning,
as described in Sec.~\ref{sec:workflow-qubo}.
We also test whether using the QUBO as part of the iterative acquisition loop
improves search performance. In this variant, the averaged-FM QUBO is solved at
every active-learning iteration and the decoded QUBO-derived candidates are
added to the evaluated set, allowing subsequent perturbations to explore the corresponding
neighborhoods.

The per-iteration-QUBO variant reaches a final best score of
$264.5\pm0.1$~GPa, compared with $264.5\pm0.1$~GPa for the reference
workflow. The trajectories overlap throughout the search, and the difference is
smaller than the observed run-to-run variation. Solving the QUBO more frequently
therefore does not improve the result on this benchmark. Together with
Sec.~\ref{sec:results-qubo-endpoint}, this shows that the QUBO is not required
to drive candidate acquisition in order for the workflow to retain a
QUBO-compatible optimization endpoint.

\section{Discussion}
\label{sec:discussion}

\subsection{Role of candidate generation and model-based selection}
\label{sec:discussion-performance}

The controlled latent-space comparisons (Sec.~\ref{sec:results-perturbation-ablation}) identify candidate generation as a major determinant of optimization performance. All latent-space methods operate through the same pretrained property-guided representation, yet their search trajectories differ substantially. Most notably, adding local perturbations to the Workflow produces a clear improvement relative to Workflow(no pert), while Random+Pert(latent) also performs strongly without a learned acquisition model. These observations indicate that exposing the search procedure to local neighborhoods around already promising latent codes is an important ingredient of effective optimization in the learned binary space. Notably, this interpretation presupposes that similar candidates are located in close proximity in latent space, underscoring the importance of jointly training the bVAE and surrogate MLP \cite{wang2020deep}. \color{black}

This result also highlights a general limitation of acquisition-based search over finite candidate pools: the acquisition function cannot select a promising candidate that is absent from the pool presented to it. In the present workflow, broad sampling maintains access to diverse regions of the binary space, whereas small bit-flip perturbations increase the density of candidates around high-performing codes. 

This distinction is supported by the variant experiments of
Sec.~\ref{sec:results-workflow-variants}. Annealing the prior--posterior mixing rate
$\alpha_{\mathrm{agg}}$ from broad exploration toward the aggregated posterior leaves the
final score unchanged within seed-to-seed variation, whereas enabling local perturbations
changes it well beyond that variation. Both are interventions on candidate generation, but
only the one that concentrates candidates near already-high-performing codes has an effect.
What matters is therefore the local density of the candidate pool rather than the global
distribution from which it is drawn.

FM-UCB then ranks candidates within this enriched pool. The experiments therefore suggest a division of roles in which candidate generation determines which regions are accessible at each iteration, while the surrogate provides model-based prioritization within those regions.

The strong performance of Random+Pert(latent) further shows that a substantial fraction of the improvement can be obtained from the local-search mechanism itself. Nevertheless, our full workflow reaches higher scores while providing a learned surrogate suitable for QUBO export. Since the two methods are not strict one-factor variants, the present benchmark does not quantify an isolated FM-UCB gain. Instead, it supports the conclusion that model-based selection provides additional utility when combined with perturbation-enhanced candidate generation.

\subsection{QUBO compatibility for data-driven optimization}
\label{sec:discussion-qubo}

For an optimization problem to be addressed by an annealing-based or other Ising optimization backends, its objective must first be expressed in a compatible binary quadratic form. In conventional combinatorial optimization this objective may be known analytically. In data-driven inverse design, however, the relationship between a candidate design and its target property is generally available only through simulations, experiments, or a predictive model. Applying a QUBO-based optimizer in this setting therefore requires an intermediate binary quadratic model that learns the objective from data. The present workflow provides such an interface. The bVAE maps the constrained alloy-design problem to binary latent variables, while the second-order FM learns the relationship between these variables and the oracle score directly as a quadratic function. In this sense, the binary representation and FM together provide a bridge between data-driven materials optimization and QUBO-based optimization.

Importantly, we find that constructing a QUBO-compatible surrogate does not require the QUBO solver itself to drive data acquisition. In the present workflow, new oracle evaluations are selected using FM-UCB over a finite candidate pool, allowing ensemble variability and local perturbations to guide the collection of training data. The final QUBO instead represents the mean quadratic surrogate and can be optimized independently of the active-learning acquisition procedure. As shown in Secs.~\ref{sec:results-qubo-endpoint} and \ref{sec:results-workflow-variants}, QUBO-derived candidates do not generally improve on the best active-learning candidates, and solving the QUBO at every iteration does not improve the final search performance. The present results therefore establish QUBO compatibility for a data-driven inverse-design workflow, rather than an advantage from quantum or annealing optimization itself. Simulated annealing is used here as a compatible QUBO solver, and the returned solutions are not assumed to be certified global optima. Demonstrating an advantage from a quantum optimization backend would require a separate comparison of solver performance and scaling and is beyond the scope of the present study.

\subsection{Limitations and outlook}
\label{sec:discussion-outlook}

\paragraph{Dependence on the latent representation.}
The optimization performance depends in part on the pretrained binary representation used by all the latent-space methods. Because this representation is property-guided, comparisons with direct composition-space optimization also reflect differences in how the search space is represented. A systematic study of alternative latent representations would help clarify the influence of representation learning on the resulting optimization performance.

\paragraph{Optimization under a machine-learned oracle.}
All optimization results are defined with respect to the pretrained \lesets{} predictor rather than directly to DFT-calculated or experimentally measured bulk modulus. The oracle provides a consistent objective for comparing optimization strategies, but an alloy that is optimal under the learned predictor need not be optimal with respect to the underlying physical property. As discussed in the supplementary information (Sec.~\ref{sec:supp-oracle-bias}), the oracle exhibits systematic prediction errors in the high-bulk-modulus regime targeted by optimization. The candidates identified here should therefore be interpreted as optima of the oracle. DFT calculations or experimental measurements of selected candidates would be required to establish whether the improvements transfer to the physical property.

\paragraph{Surrogate expressivity and future directions.}
The use of a second-order FM represents a deliberate trade-off between surrogate expressivity and direct QUBO compatibility. Higher-order factorization machines may capture more complex interactions in the latent variables~\citep{mandal2020compressed,hwang2025higher}, but would require a higher-order-to-quadratic reduction before they could be optimized with a standard QUBO backend. The present results suggest that candidate generation is at least as important as increasing surrogate complexity, but a systematic investigation of surrogate order and optimization performance remains an interesting direction for future work. More broadly, the same framework could be extended to other material properties, multiple objectives, richer compositional constraints, and other learned or simulation-based forward models, providing a route to test QUBO-compatible data-driven optimization in more demanding materials-design settings.

\section{Conclusion}
\label{sec:conclusion}

We presented a QUBO-compatible active-learning workflow for inverse design of high-entropy alloys that combines a property-guided binary latent representation with a quadratic factorization-machine surrogate. Beyond the workflow itself, the study provides a systematic benchmark of inverse-design strategies in both binary latent space and the original composition space. Controlled latent-space comparisons and ablation studies show that candidate generation is a major driver of performance: local perturbations around high-performing latent codes provide a substantial improvement, while upper-confidence-bound selection based on the factorization-machine ensemble adds model-based prioritization within the resulting candidate pool. The comparison with direct composition-space optimization further shows that the QUBO-compatible latent workflow remains competitive with strong classical search strategies, although it does not provide an optimization advantage over the best composition-space method in the present benchmark. At the same time, the learned quadratic surrogate can be exported directly as a QUBO and optimized by a compatible backend. Overall, the results establish a benchmarked route for combining data-driven materials inverse design with binary quadratic optimization while separating efficient active-learning acquisition from the choice of the final QUBO solver.
\section*{Acknowledgements}
We acknowledge the financial support of the DLR Quantum Computing Initiative through the project QuantiCoM (\url{https://qci.dlr.de/quanticom/}) funded by the German Federal Ministry of Research, Technology and Space (BMFTR).

\section*{Author contributions}
Giorgio Silvi: Methodology, Software, Investigation, Formal analysis, Writing
-- original draft, Writing -- review \& editing.
Kirsten Bark: Writing -- review \& editing.
Rolando Reiner: Methodology, Software, Investigation, Writing -- review \&
editing.
Nicolas Vogt: Supervision, Writing -- review \& editing.
Thomas Plehn: Methodology, Writing -- original draft, Writing -- review \& editing.
Daniel Barragan: Writing -- original draft, Writing -- review \& editing.
Marc Landmann: Writing -- review \& editing.
David Melching: Supervision, Writing -- original draft, Writing -- review \&
editing.
All authors contributed to the discussion and interpretation of the results and
approved the final manuscript.

\bibliographystyle{unsrtnat}
\bibliography{reference}

\clearpage
\appendix

\section{Supplementary information}
\label{sec:supplement}

\subsection{Reference dataset and GNN oracle}
\label{sec:supp-oracle}

The GNN used as an oracle in this study was originally developed and benchmarked on a DFT-calculated quaternary HEA dataset containing bulk modulus values for $N_{\mathrm{bulk}} = 7071$ compositions \citep{zhang2025graph}. Let
\begin{equation}
\mathcal{D}_{\mathrm{DFT}}
=
\left\{ (\xcomp_n, y^{\mathrm{DFT}}_n) \right\}_{n=1}^{N_{\mathrm{bulk}}}
\label{eq:supp-dft-dataset}
\end{equation}
denote the subset of this dataset with available bulk-modulus labels, where $y^{\mathrm{DFT}}_n$ is the DFT-calculated bulk modulus of composition $\xcomp_n \in \mathbb{R}^{|\mathcal{E}|}$. In the present work, however, the optimization objective is not the DFT label $y^{\mathrm{DFT}}_n$ but the prediction of the frozen GNN-oracle. We therefore construct an oracle-labeled reference dataset
\begin{equation}
\mathcal{D}_{\mathrm{oracle}}
=
\left\{ (\xcomp_n, \hat{y}_{\mathrm{GNN},n}) \right\}_{n=1}^{N_{\mathrm{bulk}}},
\qquad
\hat{y}_{\mathrm{GNN},n} = \oracle(\xcomp_n),
\label{eq:supp-oracle-dataset}
\end{equation}
so that whenever a property label is used to train or calibrate the inverse-design workflow, it is taken from the frozen oracle rather than from the original DFT target. During inverse design, newly proposed compositions are labeled by querying the same frozen oracle, $y_{\mathrm{new}} = \oracle(\xcomp_{\mathrm{new}})$. Using oracle predictions for both the reference dataset and newly proposed candidates ensures that all labels seen by the latent models, active-learning surrogates, and benchmark methods are drawn consistently from the same proxy objective. The original DFT labels are therefore not used as optimization targets in the main workflow.

\subsection{Binary VAE architecture and training objective}
\label{sec:supp-bvae-details}

The binary variational autoencoder (\bvae{}) provides the discrete latent representation used by the inverse-design workflow. The input to the encoder is not the raw composition vector, but the \textsc{LESets} global representation $\bm{Z}(\xcomp) \in \mathbb{R}^{32}$ extracted immediately before the final \textsc{LESets} prediction head. The decoder reconstructs the corresponding 15-dimensional composition vector over the fixed elemental pool. Thus, the \bvae{} learns a binary latent representation that connects the chemically-informed GNN representation to explicit alloy compositions.

For an alloy composition $\xcomp \in \mathbb{R}^{|\mathcal{E}|}$, the encoder maps the global representation $\bm{Z}(\xcomp)$ to Bernoulli logits
\begin{equation}
    \bm{a}_{\psi}(\bm{Z})
    \in
    \mathbb{R}^{B},
\end{equation}
where $B$ is the binary latent dimensionality. In the main experiments, we use $B=32$. The corresponding Bernoulli probabilities are
\begin{equation}
    \bm{p}_{\psi}(\bm{Z})
    =
    \sigma\!\left(\bm{a}_{\psi}(\bm{Z})\right),
\end{equation}
and the approximate posterior factorizes over bits:
\begin{equation}
    q_{\psi}(\zbin \mid \bm{Z})
    =
    \prod_{i=1}^{B}
    \mathrm{Bernoulli}
    \left(
        z_i;
        p_{\psi,i}(\bm{Z})
    \right).
    \label{eq:supp-bvae-posterior}
\end{equation}
During training, binary sampling is implemented using a differentiable Gumbel--Sigmoid relaxation \cite{maddison2016concrete,jang2016categorical}, and for downstream search, hard binary codes $\zbin \in \{0,1\}^{B}$ are used.

The decoder maps each latent code to real-valued composition logits
\begin{equation}
    \bm{\ell}
    =
    D_{\eta}(\zbin)
    \in
    \mathbb{R}^{15}.
\end{equation}
After applying a softmax normalization, these logits define a relaxed composition
\begin{equation}
    \widetilde{\xcomp}
    =
    \operatorname{softmax}(\bm{\ell})
    \in
    \Delta^{14},
\end{equation}
where $\Delta^d$ denotes the $d$-dimensional simplex of non-negative, normalized compositions.

For oracle evaluation, the relaxed composition is projected onto the valid quaternary composition space by retaining the four largest elemental fractions, setting all other fractions to zero, and renormalizing the retained entries.

Following the surrogate-guided generative-design strategy of Wang et al.~\citep{wang2020deep}, we train the \bvae{} together with a compact latent property surrogate
\begin{equation}
    S_{\xi} : [0,1]^B \rightarrow \mathbb{R}.
\end{equation}
During \bvae{} training, this surrogate receives the soft encoder probabilities $\bm{p}_{\psi}(\bm{Z})$ and predicts the bulk-modulus label:
\begin{equation}
    \hat{y}
    =
    S_{\xi}
    \left(
        \bm{p}_{\psi}(\bm{Z})
    \right).
\end{equation}
This training-time surrogate is used only to organize the latent space along property-relevant directions. It is distinct from the factorization-machine surrogate used later during active learning.

The total \bvae{} training objective combines a composition-reconstruction loss, a KL regularization term, and the latent property-prediction loss:
\begin{equation}
    \mathcal{L}_{\mathrm{bVAE}}
    =
    \mathcal{L}_{\mathrm{recon}}
    +
    \beta_t
    \mathrm{KL}
    \left[
    q_{\psi}(\zbin \mid \bm{Z})
    \,\|\, p_0(\zbin)
    \right]
    +
    \lambda_{\mathrm{prop}}
    \mathcal{L}_{\mathrm{prop}}.
    \label{eq:supp-bvae-total-loss}
\end{equation}
The binary latent prior is an independent fair Bernoulli distribution,
\begin{equation}
    p_0(\zbin)
    =
    \prod_{i=1}^{B}
    \mathrm{Bernoulli}(z_i;0.5).
\end{equation}
The KL term is computed exactly from the Bernoulli probabilities and is annealed during training through the weight $\beta_t$.

Because the target alloy compositions are quaternary, the reconstruction loss is
set-aware. Let $\mathcal{S}(\xcomp)$ denote the support of the four nonzero
elements in the target composition. The reconstruction loss penalizes
composition error on the true support and probability leakage outside the
support:
\begin{equation}
    \mathcal{L}_{\mathrm{recon}}
    =
    \mathcal{L}_{\mathrm{CE}}
    \left(
        \xcomp_{\mathcal{S}},
        \widetilde{\xcomp}_{\mathcal{S}}
    \right)
    +
    \lambda_{\mathrm{out}}
    \sum_{i \notin \mathcal{S}(\xcomp)}
    \widetilde{x}_i
    +
    \lambda_1
    \left\|
        \xcomp_{\mathcal{S}}
        -
        \widetilde{\xcomp}_{\mathcal{S}}
    \right\|_1.
    \label{eq:supp-set-aware-recon}
\end{equation}
The property loss is the mean-squared error between the surrogate prediction and
the standardized \lesets{} label:
\begin{equation}
    \mathcal{L}_{\mathrm{prop}}
    =
    \left\|
        S_{\xi}
        \left(
            \bm{p}_{\psi}(\bm{Z})
        \right)
        -
        y_{\mathrm{std}}
    \right\|_2^2.
    \label{eq:supp-bvae-property-loss}
\end{equation}

\begin{figure}[t]
    \centering
    \includegraphics[width=\textwidth]{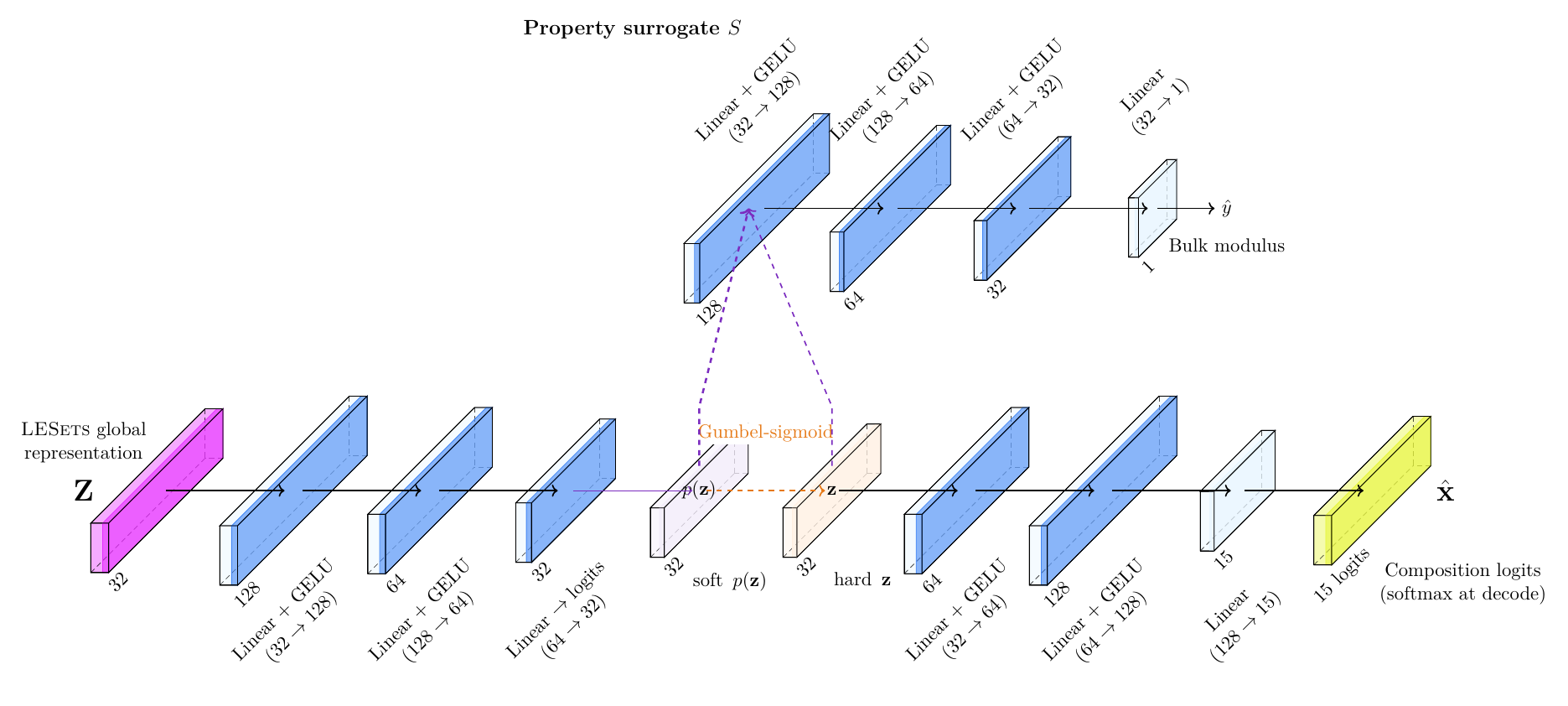}
    \caption{
    Binary VAE architecture with auxiliary latent property surrogate.
    The encoder receives the frozen \lesets{} global representation
    $\bm{Z} \in \mathbb{R}^{32}$ and produces logits for $B$ Bernoulli latent
    variables. A binary latent code is sampled using a differentiable relaxation
    during training and is passed to the decoder, which reconstructs a
    15-dimensional composition vector. In parallel, the latent surrogate
    $S_{\xi}$ predicts the frozen-\lesets{} bulk-modulus label from the soft
    encoder probabilities, encouraging property-aware organization of the
    latent space.
    }
    \label{fig:supp-bvae-architecture}
\end{figure}

Figure~\ref{fig:supp-bvae-architecture} summarizes the \bvae{} architecture and
the auxiliary latent surrogate.
% while
% Fig.~\ref{fig:supp-bvae-training-curves} shows representative training curves
% for the total loss, reconstruction loss, KL divergence, and property loss.

% \begin{figure}[t]
%     \centering
%     \includegraphics[width=\textwidth]{figures/supp_bvae_training_curves.png}
%     \caption{
%     Representative \bvae{} training curves. The total objective combines
%     composition reconstruction, KL regularization of the binary latent
%     posterior, and the auxiliary property-prediction loss. The KL weight is
%     annealed during training, allowing the model to first learn accurate
%     reconstruction and then regularize the latent distribution. 
%     \todo{Do we really need this figure? In any case, the figure is not publication-ready. The configuration at the top needs to be dropped or re-designed with the variables introduced in the paper. The training curves contain to dense information.}
%     }
%     \label{fig:supp-bvae-training-curves}
% \end{figure}
\subsection{Oracle bias and tail compression}
\label{sec:supp-oracle-bias}

The inverse-design workflow optimizes a frozen \lesets{} predictor rather than
directly optimizing reference DFT labels. To characterize systematic bias in
this oracle, we analyzed residuals over all bulk-modulus samples for
which both reference labels and \lesets{} predictions were available
($N=7071$). This diagnostic is descriptive of the oracle's bias pattern over
the available labeled dataset and is not intended as a held-out generalization
estimate.

For each composition, we define the residual as
\begin{equation}
    e_n
    =
    y^{\mathrm{DFT}}_n - \hat{y}_{\mathrm{GNN},n}.
    \label{eq:supp-oracle-residual}
\end{equation}
With this sign convention, positive residuals indicate that the oracle underpredicts the reference DFT value, whereas negative residuals indicate overprediction.

\begin{figure}[t]
    \centering
    \includegraphics[width=\textwidth]{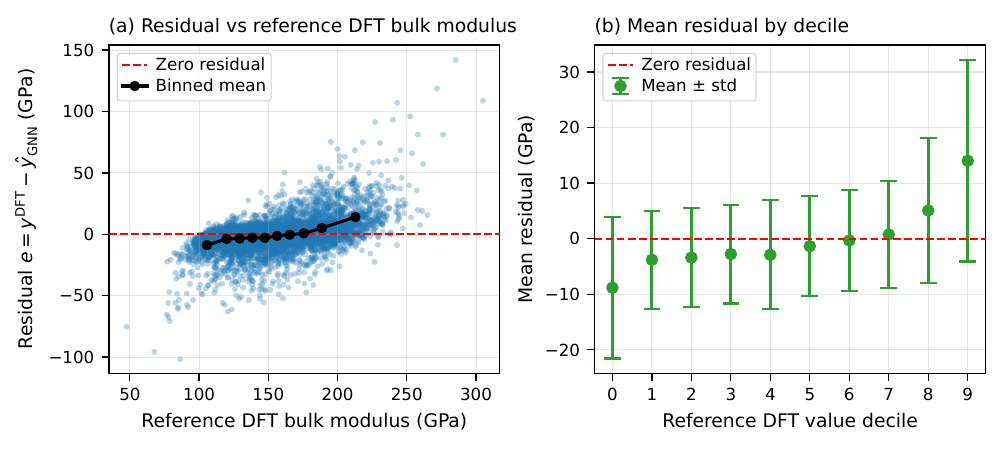}
    \caption{
    Systematic bias pattern of the \lesets{} over the
    available bulk-modulus dataset.
    \textbf{(a)} Residuals as a function of reference bulk modulus. Individual
    samples are shown as blue points, the red dashed line marks zero residual,
    and the black markers show the binned mean trend.
    \textbf{(b)} Mean residual by true-value decile, with error bars indicating
    the within-bin standard deviation. Residuals are defined as $e = y^{\mathrm{DFT}} - \oracle$, so positive values
    indicate underprediction by the oracle and negative values indicate
    overprediction. The transition from negative residuals in the lower deciles
    to positive residuals in the upper deciles indicates tail compression:
    low-bulk-modulus samples are overpredicted on average, while
    high-bulk-modulus samples are underpredicted on average. This figure is a
    descriptive diagnostic over the available labeled dataset, not a held-out
    generalization estimate.
    }
    \label{fig:supp-oracle-bias}
\end{figure}

Figure~\ref{fig:supp-oracle-bias} shows a systematic residual trend across the bulk-modulus range. Low-bulk-modulus samples tend to have negative residuals, whereas high-bulk-modulus samples tend to have positive residuals. Thus, the \lesets{} compresses the tails of the bulk-modulus distribution: low values are overpredicted and high values are underpredicted on average.

The effect is particularly relevant for inverse design targeting high bulk modulus. In the bottom 10\% of the distribution, the mean residual is $-8.83$ GPa, whereas in the top 10\% it is $+14.00$ GPa. In the more extreme 5\% tails, the mean residuals are $-12.52$ GPa and $+19.34$ GPa for the bottom and top tails, respectively. Therefore, the oracle exhibits stronger systematic bias in the high-bulk-modulus tail.

This bias does not affect the fairness of the benchmark comparisons, because all search methods are evaluated using the same oracle and the same unique-oracle-call accounting. However, it does affect the interpretation of optimized candidates. The reported optimization results should therefore be understood as improvements with respect to a fixed proxy oracle, not as direct evidence that the highest-scoring decoded alloys have the highest true DFT bulk moduli. DFT validation of top candidates would be required to assess their physical accuracy.

\subsection{Benchmark protocol and oracle-call accounting}
\label{sec:supp-benchmark-protocol}
\label{sec:benchmark-budget-cache}

The benchmark budget is defined in terms of unique oracle evaluations. A single
oracle call is counted only when a method evaluates a new unique decoded
composition. If a proposed composition has already been evaluated within the
same method run, its cached oracle value is reused and the proposal is counted
as a cache hit.

This accounting is important for latent-space methods. Different binary latent
codes can decode and project to the same quaternary composition. Counting raw
latent-code proposals would therefore overestimate the amount of new
information collected by such methods. Using unique decoded compositions as the
budget unit ensures that all methods are compared by the number of fresh
\lesets{} evaluations they obtain.

For each method and seed we track the number of raw proposals
\(N_{\mathrm{proposal}}\), the number of unique oracle evaluations
\(N_{\mathrm{unique}}\), and the number of cache hits
\(N_{\mathrm{cache}}\), with
\begin{equation}
N_{\mathrm{proposal}}
=
N_{\mathrm{unique}}
+
N_{\mathrm{cache}}.
\label{eq:supp-cache-accounting}
\end{equation}

The primary performance metric is the best oracle value found as a function of
unique oracle calls:
\begin{equation}
    y_m^{\star}(B)
    =
    \max_{
        (\xcomp_j,y_j)\in\mathcal{E}_m(B)
    }
    y_j,
\end{equation}
where $\mathcal{E}_m(B)$ is the set of unique compositions evaluated by method
$m$ up to budget $B$. Final scores are reported across independent random
seeds.

The full workflow uses a binary latent dimension $B=32$, a broad latent
candidate distribution combining the Bernoulli prior and the aggregated
encoder posterior, an ensemble of quadratic factorization-machine surrogates,
and a final QUBO verification step. The active-learning candidate pool combines
broad latent samples with local perturbations of high-performing previously
scored latent codes.

Table~\ref{tab:supp-workflow-hyperparams} summarizes the main workflow
hyperparameters. Table~\ref{tab:supp-baseline-hyperparams} summarizes the
baseline-specific settings.

\begin{table}[t]
\centering
\caption{
Main hyperparameters of the QUBO-compatible workflow.
}
\label{tab:supp-workflow-hyperparams}
\small
\begin{tabular}{ll}
\toprule
Quantity & Value \\
\midrule
Binary latent dimension & $B=32$ \\
Initial-evaluation fraction & 0.20 \\
Candidate-pool size & 20,000 \\
Aggregated-posterior fraction & $\alpha_{\mathrm{agg}}=0.5$ \\
UCB coefficient & $\beta=1.0$ \\
FM ensemble size & 5 \\
FM interaction embedding dimension & 8 \\
FM training epochs per AL iteration & 60 \\
AL batch size & 500 \\
Perturbation top-$K$ & 16 \\
Perturbation copies per top code $C$ & 64 \\
Maximum bit flips per copy $R$ & 3 \\
QUBO solver & Simulated annealing \\
QUBO reads & 20,000 \\
Final QUBO verification & Top-$k$, $k=5$ \\
\bottomrule
\end{tabular}
\end{table}

\begin{table}[t]
\centering
\caption{
Baseline-specific hyperparameters used in the benchmark.
}
\label{tab:supp-baseline-hyperparams}
\small
\begin{tabular}{lll}
\toprule
Method & Hyperparameter & Value \\
\midrule
Random(comp) & Element support & 4 elements sampled uniformly \\
Random(comp) & Fractions & Dirichlet$(1,1,1,1)$ \\
GA(comp) & Population size & 128 \\
GA(comp) & Parent selection & Tournament selection \\
GA(comp) & Crossover & Convex composition blending \\
GA(comp) & Mutation & Element swap and multiplicative fraction noise \\
GA(comp) & Constraint handling & Repair to exactly four nonzero fractions \\
RF-UCB(comp) & Surrogate & Random forest \\
RF-UCB(comp) & Number of trees & 200 \\
RF-UCB(comp) & Minimum leaf size & 2 \\
RF-UCB(comp) & Acquisition & $\mu+\beta\sigma$ \\
RF-UCB(comp) & Candidate pool per step & 20,000 fresh valid compositions \\
Random(latent) & Sampling distribution & Prior/aggregated-posterior mixture \\
GA(latent) & Population size & 128 \\
GA(latent) & Crossover & Uniform bit crossover \\
GA(latent) & Mutation & Bit flips, at least one bit flipped \\
Random+Pert(latent) & Warm-start fraction & 0.15 of budget, minimum 50 proposals \\
Random+Pert(latent) & Iteration mixture & 80\% broad samples, 20\% perturbations \\
Random+Pert(latent) & Perturbation source & Top scored latent codes so far \\
\bottomrule
\end{tabular}
\end{table}

\end{document}